\documentclass[superscriptaddress,showpacs,preprintnumbers,amsmath,amssymb,twocolumn,prb]{revtex4-1}
\usepackage{graphicx}
\usepackage{dcolumn}
\usepackage{bm}
\usepackage{color}
\usepackage{epstopdf}
\renewcommand{\bm}{\mathbf}

\usepackage[dvipsnames]{xcolor}

\begin{document}


\title{Spin-dependent Raman and Brillouin light scattering on excitons \\ in CsPbBr$_3$ perovskite crystals 
}

\author{Ina V. Kalitukha}
\affiliation{Ioffe  Institute, Russian Academy of Sciences, 194021 St. Petersburg, Russia}
\author{Victor F. Sapega}
\affiliation{Ioffe  Institute, Russian Academy of Sciences, 194021 St. Petersburg, Russia}
\author{Dmitri R. Yakovlev}
\affiliation{Experimentelle Physik 2, Technische Universit\"{a}t Dortmund, 44227 Dortmund, Germany} 
\affiliation{Ioffe Institute, Russian Academy of Sciences, 194021 St. Petersburg, Russia}
\author{Dennis Kudlacik}
\affiliation{Experimentelle Physik 2, Technische Universit\"{a}t Dortmund, 44227 Dortmund, Germany}
\author{Damien Canneson}
\affiliation{Experimentelle Physik 2, Technische Universit\"{a}t Dortmund, 44227 Dortmund, Germany}
\author{Yury G. Kusrayev}
\affiliation{Ioffe  Institute, Russian Academy of Sciences, 194021 St. Petersburg, Russia}
\author{Anna V. Rodina}
\affiliation{Ioffe  Institute, Russian Academy of Sciences, 194021 St. Petersburg, Russia}
\author{Manfred Bayer}
\affiliation{Experimentelle Physik 2, Technische Universit\"{a}t Dortmund, 44227 Dortmund, Germany} 
\affiliation{Research Center FEMS, Technische Universit\"at Dortmund, 44227 Dortmund, Germany}

\date{\today}

\begin{abstract}
The spin properties of excitons and charge carriers  in CsPbBr$_3$ lead halide perovskite crystals are investigated by spin-dependent light scattering in magnetic fields up to 10~T. Spin-flip Raman scattering spectra measured under resonant excitation of exciton-polaritons show a rich variety of features provided by the Zeeman splittings of excitons and of electrons and holes interacting with the excitons. The magnitudes and anisotropies of their Land\'e $g$-factors are measured. A detailed consideration of the responsible mechanisms is presented and discussed in relation to the experimental data, in particular on the polarization properties of the Raman spectra. 
We consider several mechanisms for the combined spin-flip Raman scattering processes involving resident carriers and photoexcited excitons and suggest new ones, involving trions in the intermediate scattering state. A double electron spin-flip caused by the exciton interaction with two localized or donor-bound electrons is revealed. The spectral lines of Brillouin light scattering on  exciton-polaritons shift in energy and become polarization-sensitive in magnetic field, evidencing the splitting of the exciton-polariton dispersion. 
\end{abstract}

\maketitle

Correspondence and requests for materials should be addressed to I.V.K. (email: kalitukha@gmail.com) and D.R.Y. (email: dmitri.yakovlev@tu-dortmund.de)


\section{Introduction}

Lead halide perovskite semiconductors attract great attention due to their remarkable photovoltaic efficiency~\cite{nrel2025} and optoelectronic properties~\cite{Vinattieri2021_book,Vardeny2022_book}, being also promising candidates for spintronics applications~\cite{Vardeny2022_book,wang2019,kim2021}. All-inorganic perovskites with Cs as anion have advanced stability at ambient conditions compared with hybrid organic-inorganic materials. Their single crystals with good structural and optical properties can be synthesized in a solution~\cite{Stoumpos2013,Dirin2016,nazarenko2017} or by Bridgman growth from a melt~\cite{Nitsch1996,He2018,Zhang2018}. CsPbBr$_3$ is a material representative for the lead halide perovskite semiconductors, which initial optical studies date back to 1978~\cite{Heidrich78,Ito78,Froelich1979,Froelich1979b,Heidrich1981}. It has been used for fabrication of solar cells~\cite{Kulbak2015,Kulbak2016}, sensitive visible light detectors~\cite{Song2016} and high-energy detectors~\cite{Stoumpos2013,Matt2020}. In applications for hard radiation detection, CsPbBr$_3$ is indispensable as it is the only lead halide perovskite that can operate under high bias as required for efficient extraction of charge carriers~\cite{Stoumpos2013}.
 
The optical properties of semiconductors in the vicinity of the band gap energy are determined by excitons, making knowledge of the exciton parameters of great importance for optoelectronics. In CsPbBr$_3$ the exciton binding energy reaches 32.5~meV~\cite{Yang2017,Yakovlev2023_High_B}, and the light-matter interaction is quite strong~\cite{su2020,Su2021,tao2022,Dursun2018}, which results in a pronounced resonance in reflectivity and absorption spectra in the vicinity of the band gap, showing the features of exciton-polaritons~\cite{Klingshirn_book_2005}. The interaction can be further enhanced in a microcavity~\cite{Bao2019,su2020}.  At present, the available information on the exciton level structure and the exciton-polariton dispersion as well as their modifications in magnetic field is limited.

Among the spin-dependent properties of CsPbBr$_3$ including the parameters that control them, the electron and hole Land\'e factors ($g$-factors) were measured by time-resolved Kerr rotation~\cite{belykh2019} and spin-flip Raman scattering~\cite{kirstein2022nc}.   Their anisotropy is provided by the orthorhombic structural phase of CsPbBr$_3$ crystals. Note that the $g$-factor values and anisotropy give access to the band structure parameters, that also enter the charge carrier effective masses~\cite{Yu2016,kirstein2022nc}. The exciton $g$-factor was measured through the Zeeman splitting of the exciton resonance in reflectivity in high magnetic fields up to 60~T~\cite{kopteva2023_gX}. Carrier spin relaxation times up to 50~ns were reported at cryogenic temperatures for localized carriers, which spin dynamics were limited by their interaction with the nuclear spin fluctuations~\cite{belykh2019}. 

Light scattering provides powerful optical techniques to study the charge and spin states in semiconductors~\cite{Yu_Cardona_book_1996, Klingshirn_book_2005}. In case of resonant excitation of exciton states, the light scattering may involve acoustic or optical phonons as well as charge and spin excitations, which allows one to get detailed information on the exciton energy and spin level structure and on the charge carriers interacting with the excitons. Brillouin light scattering allows one to reconstruct the energy dispersion of exciton-polaritons~\cite{Ulbrich1977,Koteles1979,Honerlage1985}. Applying a strong magnetic field, which is an established tool of exciton spectroscopy, provides access to spin-dependent properties and phenomena. 

Spin-flip Raman scattering (SFRS) is a reliable technique for measuring exchange and Zeeman splittings and thus evaluating carrier and exciton $g$-factors. It has a high spectral selectivity in excitation that can be combined with spectral resolution, and exploits strict selection rules by analyzing the linear and circular light polarization. SFRS can be resonantly enhanced for excitation on the exciton states, and often provides straight a forward interpretation of the measured data. The potential of the technique has been demonstrated for bulk materials~\cite{Hopfield1968, Scott1972}, as well as for semiconductor nanostructures, namely quantum wells~\cite{Sapega1992, Sapega1994,Sirenko1997} and quantum dots\cite{Sirenko1998,Debus2014,Kudlacik2019}. Recently, SFRS was used to measure the electron and hole $g$-factors in the lead halide perovskite semiconductors~\cite{kirstein2022nc,Harkort_2D_2023}. Theoretically, the mechanisms of SFRS in perovskite semiconductors were considered in Refs.~\cite{Rodina2022,Rodina2024}.

In this paper, we investigate CsPbBr$_3$ perovskite crystals by means of the spin-flip Raman scattering and Brillouin scattering techniques. Resonant excitation of the exciton states is used. The signals are measured in close vicinity of the laser line (spectral shifts in the range from 0.1 to 3~meV), at cryogenic temperatures in strong magnetic fields up to 10~T. Spin-flip lines of charge carriers and excitons are observed and electron, hole and exciton $g$-factors as well as their anisotropies are measured. We discuss the underlying mechanisms and show that the exciton serves as a mediator for the carrier spin-flip scattering. An unusual double electron spin-flip is found and assigned to the exciton interaction with two localized or donor-bound electrons. In addition, spin-dependent Brillouin light scattering on the exciton-polaritons is found. 


\section{Experiment}

\textbf{Samples.} 
We study a preselected, solution-grown CsPbBr$_3$ perovskite single crystal. Its high structural quality allows us to minimize detrimental effects related to inhomogeneous broadening of the exciton resonances. The crystal has an elongated shape with sizes of $5\times 2 \times 2$~mm$^3$. The longer direction corresponds to the $c$-axis [direction (001)], as shown in the sketch in Figs.~\ref{fig:1}(b) and \ref{fig:SI1}, the $a$ and $b$-axes are equivalent in our experiments. 

\textbf{Photoluminescence and reflectivity.} 
Photoluminescence (PL) was excited by a continuous-wave Ar-ion laser with the photon energy of 2.540~eV (wavelength of 488~nm). a low excitation density of 1.5~Wcm$^{-2}$ was used. The emission was analyzed by a 0.5-meter spectrometer and a liquid-nitrogen-cooled charge-coupled-device. This setting was used for reflectivity measurements. The reflectivity spectrum was recorded in back-scattering geometry, using a plasma-lamp providing a continuous spectrum. The sample was kept in pumped liquid helium at the temperature of $T=1.6$~K.

\textbf{Spin-flip Raman scattering (SFRS).} 
Resonant Raman spectra were measured at $T=1.6$~K in backscattering geometry with 2.3305~eV photon energy (wavelength of 532~nm) and an incident power in the range of $1-5$~Wcm$^{-2}$. The scattered light was collected within a solid angle 0.03 sr ($\approx6$ deg in a plane) and analyzed by a Jobin-Yvon U1000 double monochromator equipped with a cooled GaAs photomultiplier interfaced with conventional photon counting electronics. The width of the monochromator slits was chosen to provide a spectral resolution of 0.2~cm$^{-1}$ (0.024~meV) or 0.5~cm$^{-1}$ (0.060~meV), which allows us to measure Raman signals in close vicinity of the laser line for spectral shifts in the range from 0.1 to 3~meV.  For  most of the presented experiments the sample was mounted in a strain-free holder and was in direct contact with liquid helium. External magnetic fields up to 10~T were generated by a superconducting split-coil solenoid.

The Raman spectra were taken in co- and cross- circular polarization in the Faraday geometry ($\mathbf{B}_{\rm F} \parallel \mathbf{k}$, $\theta=0^{\circ}$), as well as in co- and cross- linear polarization in the Voigt geometry ($\mathbf{B}_{\rm V} \perp \mathbf{k}$, $\theta=90^{\circ}$), where $\bf k$ is the incident light wave vector and $\theta$ is the angle between the magnetic field $\textbf{B}$ and $\textbf{k}$ directions. The polarization configuration for Raman spectra is denoted as $ab$, where $a$ is the polarization of the incident light ($\sigma^+$ or $\sigma^-$ in circular basis, horizontal (H) or vertical (V) in linear basis), and $b$ is the detected light polarization. In most measurements the sample c-axis was either perpendicular to the light $k$-vector ($\textbf{c} \perp \textbf{k}$, $\theta_k = 90^{\circ}$), or parallel to it ($\textbf{c} \parallel \textbf{k}$, $\theta_k = 0^{\circ}$), where $\theta_k$ is the angle between the c-axis and $\textbf{k}$.
In order to measure different components of the $g$-factor tensor, the angle $\theta_B$ between the magnetic field $\textbf{B}$ and the c-axis was varied in the range of $0^{\circ} \le \theta_B \le 90^{\circ}$ by rotating the sample. During this process, the angle between magnetic field and incident laser remained the same for both geometries, namely $\mathbf{B}_{\rm V} \perp \mathbf{k}$ for the  Voigt geometry while tilting the sample ($0^{\circ} \le \theta_B \le 45^{\circ}$) and $\mathbf{B}_{\rm F} \parallel \mathbf{k}$ for the Faraday geometry while tilting the sample ($45^{\circ} \le \theta_B \le 90^{\circ}$). The schemes of all used geometries are shown in Fig.~\ref{fig:SI8} in SI.

In our experiment the excitation laser had a small but finite incidence angle $15^{\circ}$ with the Faraday magnetic field $\mathbf{B}_{\rm F}$ (and correspondingly $\theta=85^{\circ}$ between ${\bm k}$ and $\mathbf{B}_{\rm V}$). Due to the difference between the refractive indices of liquid helium (1.05) and CsPbBr$_3$ (1.5)~\cite{Ahmad2017} the photon wave vector angle reduces to about $10^{\circ}$ in the crystal. This means that the Faraday/Voigt magnetic fields are not strictly parallel/perpendicular to the incident phonon wave vector. For simplicity, we give the $\theta$ and $\theta_k$ values still as $0^{\circ}$ or $90^{\circ}$ in the experimental section. The contribution of this small deviation of the incident angle manifests itself in making the spin-flip of a single electron allowed in experiments in the Faraday geometry, while it is forbidden by theory in the ideal Faraday geometry.

\section{Reflectivity and photoluminescence spectra}

\begin{figure*}
\centering
\includegraphics[width=1\textwidth]{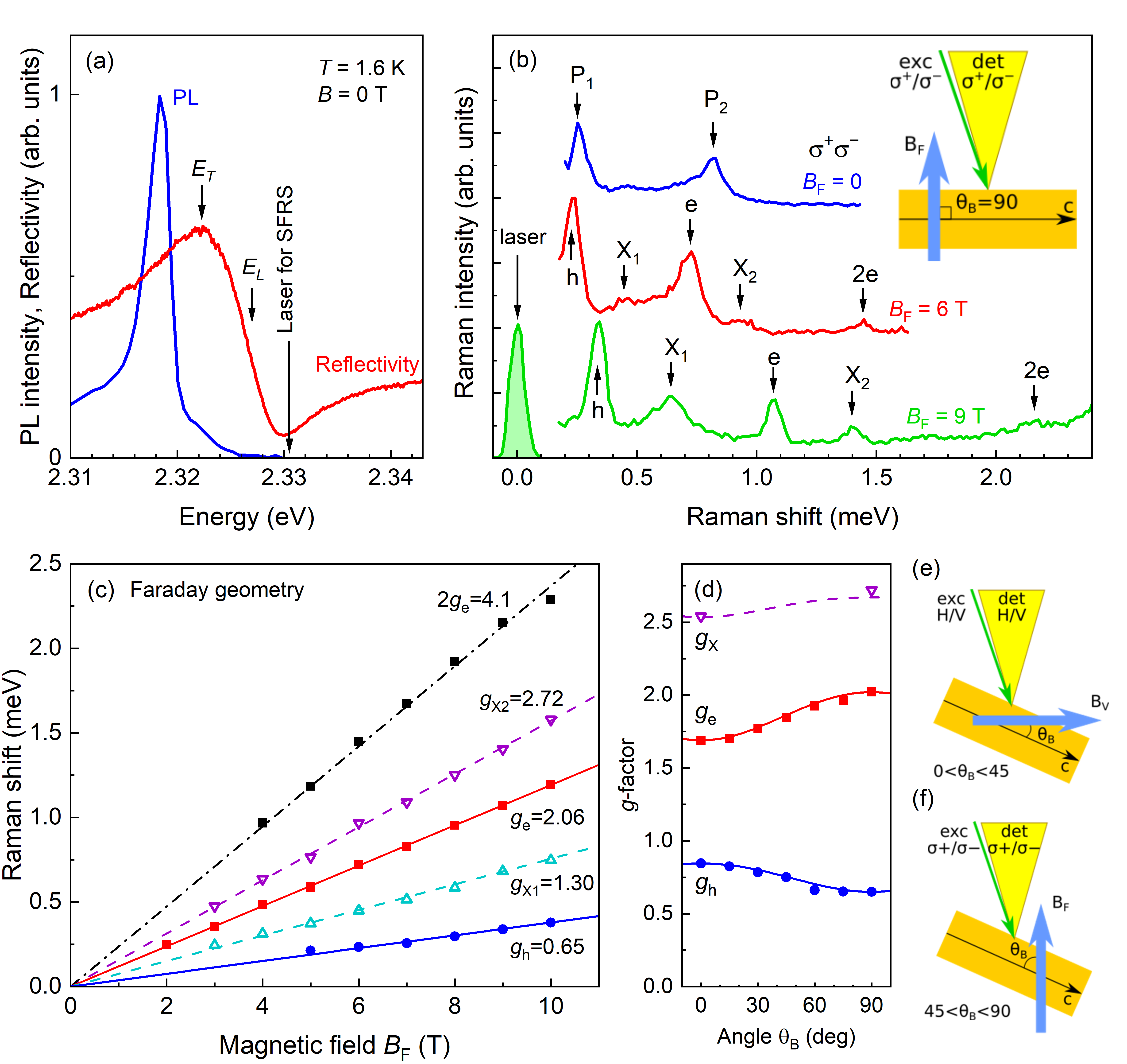}
\caption{{Raman spectra of the CsPbBr$_3$ crystal in magnetic field.} 
 ({a}) Reflectivity (red line) and photoluminescence (blue line) spectra measured at $T=1.6$~K. The  PL is excited at 2.540~eV. The rrow at 2.3305~eV indicates the laser energy used for SFRS measurements.  
({b}) Light scattering in cross circular polarization configuration $\sigma^+ \sigma^-$: Brillouin scattering spectrum on exciton-polaritons at zero magnetic field (blue line). SFRS spectra at $B_{\rm F}=6$~T and 9~T in Faraday geometry ($\mathbf{B}_{\rm F} \parallel \mathbf{k}$ and  $\mathbf{B} \perp \mathbf{c}$, as shown in inset). The Stokes shifted signals correspond to positive Raman shifts. The hole (h), electron (e), and exciton (X$_1$, X$_2$) spin-flip lines are clearly seen. Also, a double electron (2e) spin-flip line can be resolved.  
({c}) Faraday magnetic field dependences of the Raman shifts of the electron, hole, exciton and double electron SFRS lines. The experimental data are shown by the symbols. The lines are linear fits using Eq.~\eqref{eq:Zeeman} to evaluate the $g$-factors. The solid lines for the charge carriers give $g_{e\perp}=2.06$ and $g_{h\perp}=0.65$. The dashed line for the X$_2$ exciton gives a fit with $g_{X2\perp}=2.72$, which exactly matches the relation $g_X=g_e+g_h$. 
({d}) Angular dependence of the electron, hole, and exciton $g$-factors. Here the $g_X$ corresponds to the $g_{X2}$ from panel (c). The solid lines are fits with Eq.~\eqref{eq:anisotropy}.
({e, f}) Scheme of the experimental geometry for the sample tilted in the Voigt (e) and the Faraday (f) geometries.
} 
\label{fig:1}
\end{figure*}

The exciton resonance in CsPbBr$_3$ contributes very pronouncedly in the reflectivity and photoluminescence spectra measured at $T=1.6$~K temperature. One can see in Fig.~\ref{fig:1}(a) that the reflectivity spectrum shows a strong exciton-polariton resonance with the typical derivative shape. Modeling with the formalism of Ref.~\onlinecite{Hopfield1963} allows us to get the energies of the longitudinal ($E_L=2.3273$~eV) and transverse ($E_T=2.3220$~eV) exciton-polaritons, and the longitudinal-transverse splitting of $\hbar \omega_{LT}=5.3$~meV, see Ref.~\onlinecite{Yakovlev2023_High_B}. The latter value characterizes the exciton oscillator strength and mquantifies the light-matter interaction at the exciton-polariton resonance. The PL spectrum has a narrow line with a maximum at 2.318~eV  and a full width at half maximum of 2.4~meV, evidencing the high structural and optical  quality of the studied CsPbBr$_3$ crystal. This PL line is  Stokes shifted from $E_T$ by 4 meV, which can be assigned to recomibination of either bound excitons or spatially separated electrons and holes~\cite{kirstein2022am,kirstein2022_mapi}, while its nature needs further clarifications. The high energy shoulder of this line with maximum at about 2.322~eV, which coincides with $E_T$, can be assigned to emission of free excitons.

\section{Spin-flip Raman scattering}

Resonant Raman spectra are measured at the laser energy of 2.3305~eV, which is 
close to the resonance of the free exciton energy at $T=1.6$~K, see the arrow in Fig.~\ref{fig:1}(a). The spectrum measured for crossed circular polarizations ($\sigma^+$ excitation and $\sigma^-$ detection) at $B=0$~T is shown in Fig.~\ref{fig:1}(b). Here, zero Raman shift corresponds to the laser energy and positive shifts correspond to Stokes-shifted lines, i.e. shifted to lowrt energies relative to the laser. The spectrum has two lines marked P$_1$ and P$_2$. These lines originate from Brillouin light scattering on exciton-polaritons and will be discussed in Section~\ref{brill}.

Two other Raman spectra in Fig.~\ref{fig:1}(b)  are measured in a magnetic field applied in Faraday geometry so that $\mathbf{B}_{\rm F} \parallel \mathbf{k}$ and  $\mathbf{B} \perp \mathbf{c}$, see the inset. One can see that in a Faraday magnetic field ($B_{\rm F} = 6$ and $9$~T) light scattering on exciton-polaritons (i.e. lines P$_1$ and P$_2$)  is suppressed, but several new lines appear. They shift linearly from the laser photon energy with increasing field strength, i.e., have characteristic features of spin-flip lines for charge carriers and excitons as reported for II-VI and III-V semiconductors~\cite{Sapega1992,Sirenko1997}. These lines are associated with spin-flips of hole (h), electron (e), and double electron (2e), discussed in Section~\ref{carriers}, and with exciton spin-flips (X$_1$ and X$_2$), discussed in Section~\ref{exc}. 

The linear shift of the spin-flip lines with increasing magnetic field can be directly associated with the Zeeman splitting of the respective electronic states
\begin{equation}
    \Delta E=|g|\mu_{\rm B}B \, ,
\label{eq:Zeeman}
\end{equation}
where $g$ is the Land\'e $g$-factor and $\mu_{\rm B}$ is the Bohr magneton. The fan plot for these lines as function of the magnetic field is given in Fig.~\ref{fig:1}(c) along with the corresponding $g$-factor values. 

\subsection{Spin-flip Raman scattering on carriers} \label{carriers}

Let us first discuss the properties of the spin-flip lines, which we assign to the hole having a smaller Raman shift than the electron showing a larger shift. Their $g$-factors perpendicular to the c-axis, measured in the Faraday geometry, are $g_{h\perp}=0.65$ and $g_{e\perp}=2.06$, respectively. We make this line assignment based on the $g$-factor values according to the universal dependence of the electron and hole $g$-factors on the band gap energy~\cite{kirstein2022nc}. We showed in this paper that both the electron and hole $g$-factors are positive in CsPbBr$_3$. Note that the $g$-factor sign cannot directly be identified from the SFRS spectra.  

As one can see in Fig.~\ref{fig:1}(c), Raman shifted-lines for the elecron and the hole can be detected when they exceed about 0.2~meV. The linear extrapolation of the shifts to zero magnetic field with high accuracy gives zero shift values. Also, we measure the anisotropy of the $g$-factors  by rotating the sample c-axis with respect to the magnetic field direction, see Fig.~\ref{fig:1}(d). Changing the angle $\theta_B$ from $90^\circ$ ($\textbf{B} \perp \textbf{c}$, Faraday geometry) to $0^\circ$ ($\textbf{B} \parallel \textbf{c}$, Voigt geometry)  results in an increase of the hole $g$-factor to $g_{h\parallel}=0.85$
and a decrease of the electron $g$-factor to $g_{e\parallel}=1.69$, see also Table~\ref{tab}. The angle dependences of the $g$-factors can be well fitted with
\begin{equation}
    g_{e(h)}= \sqrt{g^2_{e(h)\parallel} \cos^2 \theta_B +g^2_{e(h)\perp} \sin^2\theta_B } \, ,
\label{eq:anisotropy}
\end{equation}
see the solid lines in Fig.~\ref{fig:1}(d). 

\begin{table}[h!]
\centering
\begin{tabular}{|c|c|c|}
\hline
& \shortstack{ $g_\perp$ \\ $\mathbf{B} \perp \mathbf{c}$  }  & \shortstack{ $g_\parallel$ \\ $\mathbf{B} \parallel \mathbf{c}$  } \\ \hline
hole $g_h$ & 0.65 & 0.85 \\ \hline
electron $g_e$ & 2.06 & 1.69 \\ \hline
exciton $g_X$ & 2.72 & 2.54 \\ \hline
\end{tabular}
\caption{Experimentally measured $g$-factors of charge carriers and excitons in a CsPbBr$_3$ crystal.}
\label{tab}
\end{table}

It is interesting to note that in the Raman spectra shown in Fig.~\ref{fig:1}(b) one can identify a weak line, which shift in magnetic field is twice larger than the electron Zeeman splitting, corresponding to $2g_{e}=4.1$, see Fig.~\ref{fig:1}(c).  This line is more pronounced in crossed circular polarizations. We assign it to a double electron spin-flip process, where the exciton interacts with two electrons of the same orientation within its localization volume and both electrons flip their spins. This is a rarely observed phenomenon, which has been previously reported for bulk CdS~\cite{Scott1972, Economou1972}, ZnTe~\cite{Oka1981}, and CdSe colloidal nanoplatelets~\cite{Kudlacik2019}.

\begin{figure}[h!]
\centering
\includegraphics[width=0.5\textwidth]{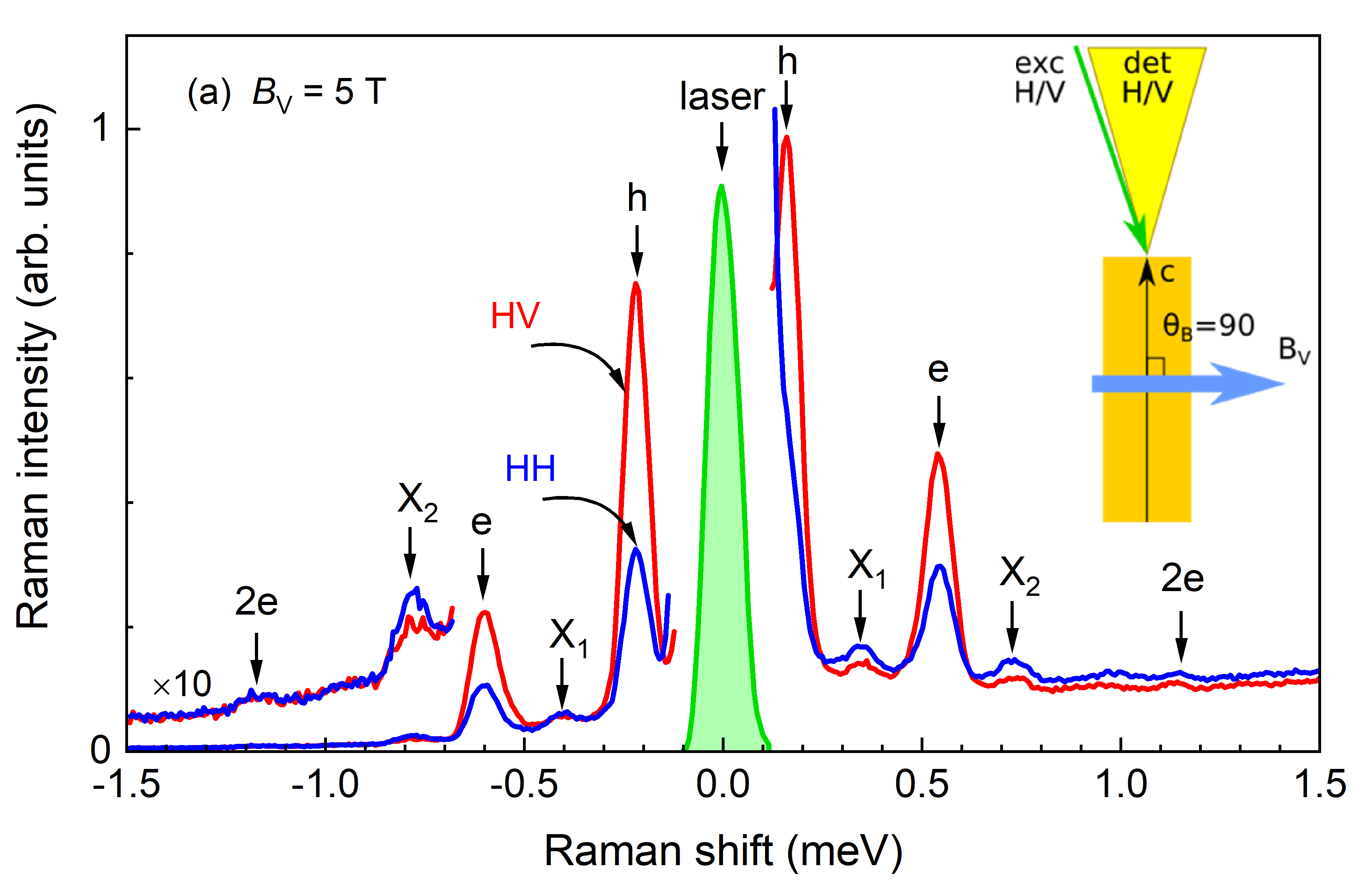}
\includegraphics[width=0.45\textwidth]{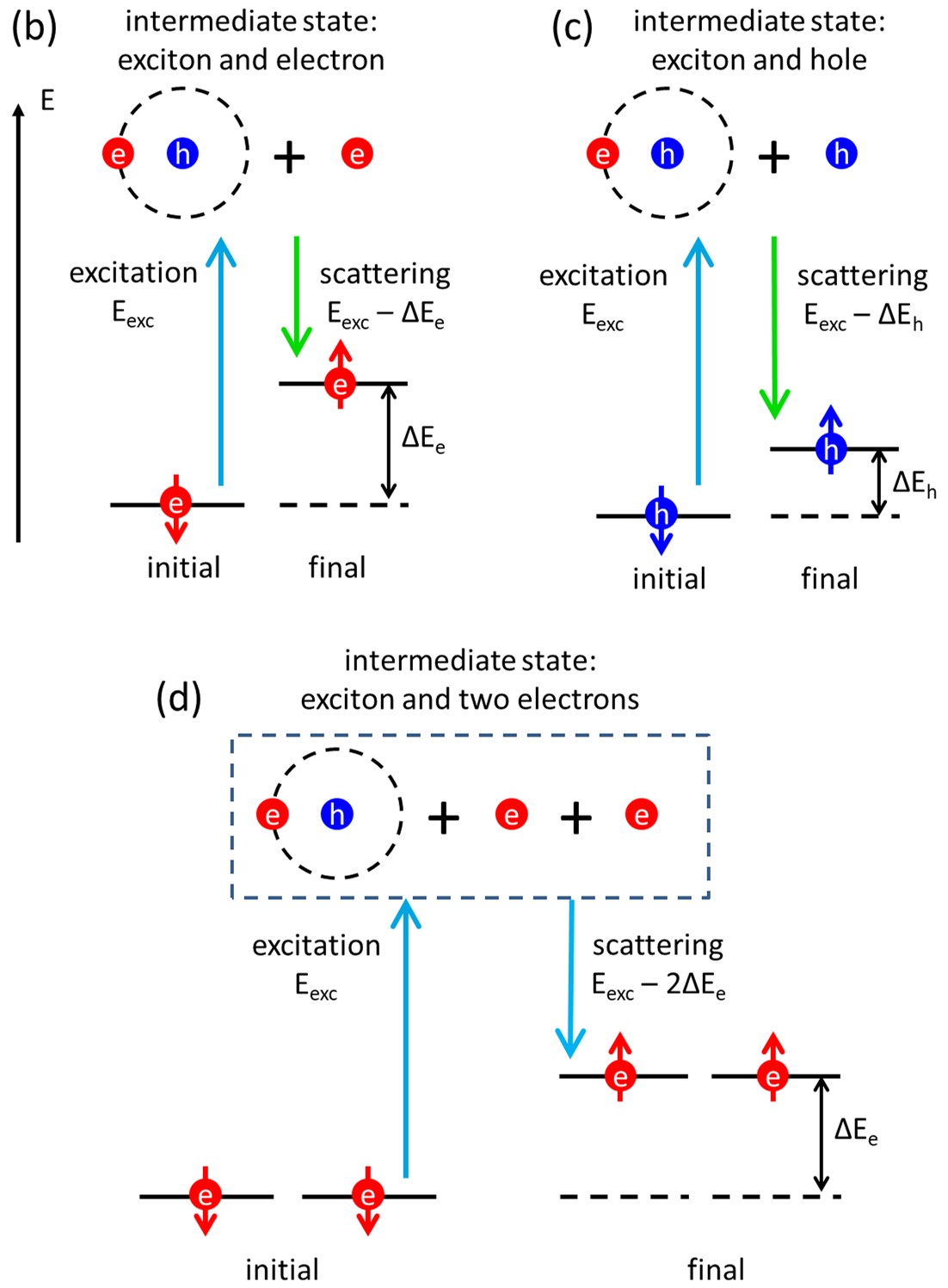}
\caption{{SFRS spectra in Voigt geometry.} (a) Spin-flip Raman scattering spectra measured at $B_{\rm V}=5$~T using co- (HH, blue line) and cross- (HV, red line) linear polarizations. Laser photon energy is 2.3305~eV, $T=1.6$~K, and $\mathbf{B}_{\rm V} \perp \mathbf{c}$. (b, c, d) Mechanisms of carrier spin flips for electron (b), hole (c), and double electron (d). The different colors of excitation (blue) and detection (green) arrows in panels (b,c) show that the signal is obtained in cross-linear polarized configuration (HV or VH), while using the same colors in panel (d) indicates processes active in co-linear polarized configurations (HH or VV).
} 
\label{fig:S2}
\end{figure}

Figure~\ref{fig:S2}(a) shows the polarization properties of the spin-flip lines measured in co- and cross-linear polarized  configurations in the magnetic field of $B_{\rm V}=5$~T applied in the Voigt geometry ($\mathbf B_{\rm V} \perp \mathbf k$, $\mathbf B_{\rm V } \perp \mathbf c$). The linear polarizations are denoted as H for horizontal polarization parallel to $\mathbf B_{\rm V }$ and V for vertical polarization orthogonal to $\mathbf B_{\rm V }$. The Brillouin light scattering on exciton-polaritons (lines P$_1$ and P$_2$ in Fig.~\ref{fig:1}(b)) is strongly suppressed in this experimental configuration as shown in Fig.~\ref{fig:S2}(a). This makes the polarization properties of the spin-flip lines more transparent in both the Stokes and anti-Stokes ranges of the spectra. Note that the spin-flip lines of electrons and holes are more prominent in the cross-polarized configuration. The double electron spin-flip line (2e) is detected with weak polarization dependence. 

\subsection{Mechanisms of electron and hole  spin-flip and polarization selection rules}
\label{mecanism_eh}

The single electron spin-flip (e-SF) line observed in Fig.~\ref{fig:S2}(a) in the Voigt geometry can be explained as the spin-flip of an electron in a three-particle complex. This complex is composed of two electrons and one hole and is contributed by a resident localized electron (using this term we also mean a donor-bound electron) and the photogenerated exciton. 

The electron spin-flip process is shown schematically in Fig.~\ref{fig:S2}(b). Before photoexcitation the resident electron occupies the lowest Zeeman state with spin down. The incident laser photon with energy $E_{\rm exc}$ creates an exciton interacting with the resident electron. Then, within the coherence time, the resident electron flips its spin due to direct or indirect exchange interaction, see Ref. \onlinecite{Rodina2022} for a detailed description, with the photocreated electron. Then, for the scattered light, the photon energy is reduced to $E_{\rm exc} - \Delta E_e = E_{\rm exc} - g_e\mu_BB$ and the resident electron is left with spin up orientation. Such processes were observed for the donor-bound electron spin-flip in bulk semiconductors~\cite{Scott1972} and for the resident electron spin-flip in semiconductor nanostructures~\cite{Kudlacik2019}. An equivalent consideration can be applied for the spin-flip of the resident hole involving the intermediate "exciton + hole" state with $E_{\rm exc} - \Delta E_h = E_{\rm exc} - g_h\mu_BB$, see Fig.~\ref{fig:S2}(c). 

\begin{figure*}[!htb]
\centering
\includegraphics[width=0.9\textwidth]{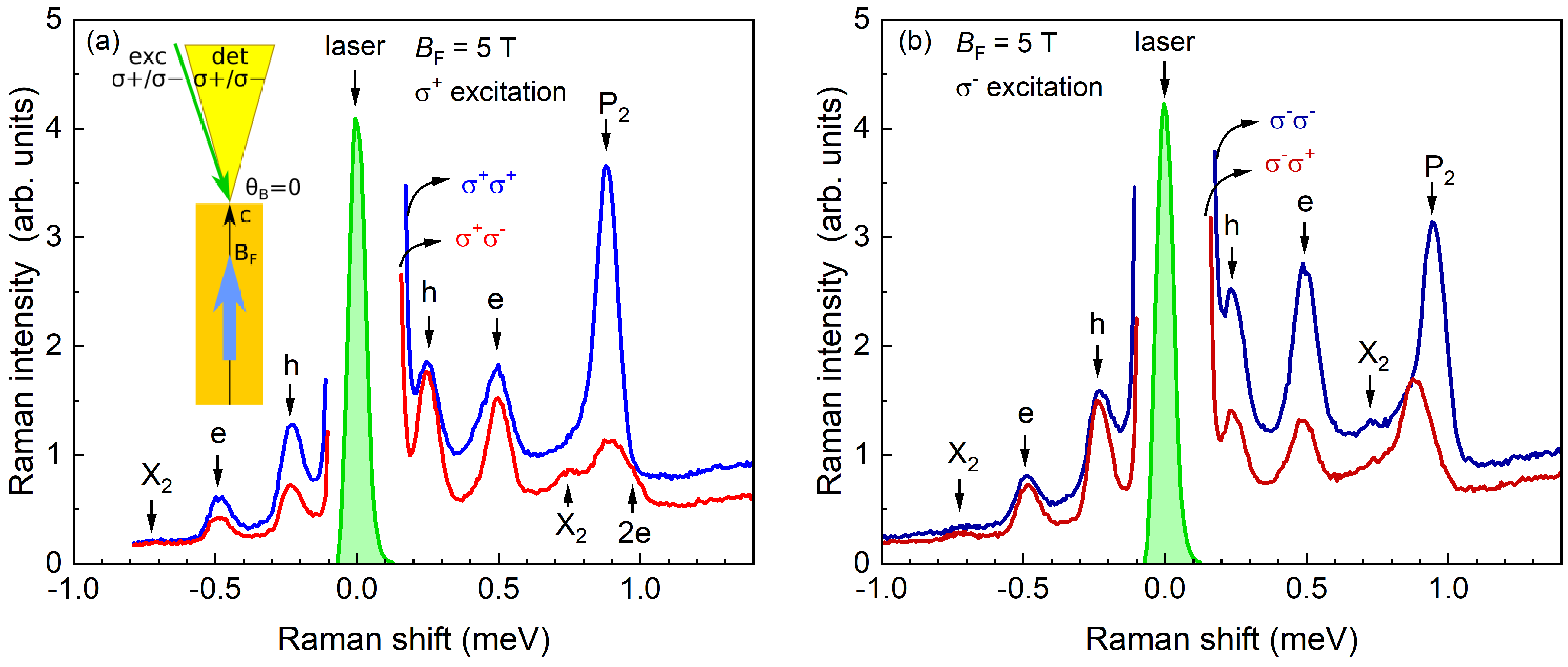}
\includegraphics[width=0.6\textwidth]{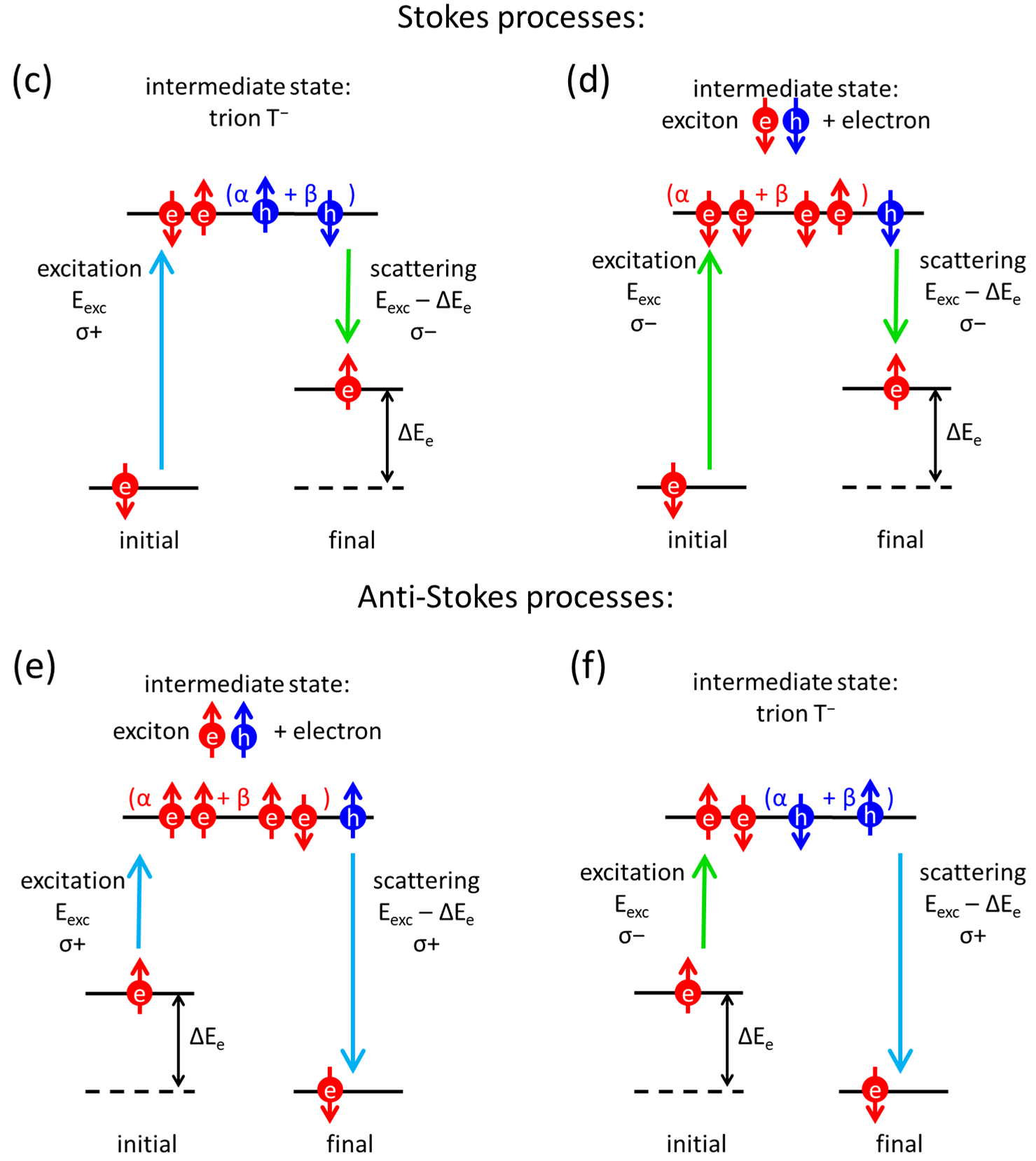}
\caption{(a), (b) Raman spectra in the Faraday geometry for all 4 configurations of excitation-detection circular polarization. The experimental geometry is shown in the inset of (a): the magnetic field is parallel to the sample c-axis, $\theta_B = 0^\circ$. The polarization properties almost coincide with the case of a Faraday magnetic field oriented perpendicular to the crystal c-axis (see Fig.~\ref{fig:SI4}). The Brillouin light scattering lines are relatively strong for the shown configuration. (c-f)  Schemes of mechanisms of electron spin-flip. Note, that the mechanisms for hole spin-flip work analogously, and therefore are not shown. The colors of the excitation and detection arrows stand for their polarizations: blue for $\sigma^+$ and green for $\sigma^-$.  }
\label{fig:SI5}
\end{figure*}

We remind here that in the perovskite semiconductors both electrons and holes have spin 1/2. In magnetic field, their initial and final states can be characterized by their projections on the field, while in the photogenerated exciton their projection on the wave vector direction $\bm k$ can be used. By direct and indirect exchange interaction between the resident and photogenerated carriers we denominate two possible ways of the resident carrier spin-flip connected with direct and indirect pathways for the recombination of excitons, respectively \cite{Rodina2020,Rodina2022}. In the indirect recombination channel, the electron (or hole) in the exciton recombines with the resident hole (or electron) and the remaining photoexcited charge carrier takes the place of the resident electron (hole), but with opposite spin. In the direct recombination channel of the photogenerated carriers, beforehnd the resident carrier spin-flip takes place, mediated by the direct interaction $ \propto  {\bm \sigma}_{e(h)}^{\rm r} \cdot {\bm \sigma}_{e(h)}$. Here $ {\bm \sigma}_{e(h)}^{\rm r}$ and ${\bm \sigma}_{e(h)}$ are the three-component pseudovectors composed of the Pauli matrices acting on the spin states of the resident and photoexcited electrons (holes), respectively. 

In the cubic phase of the perovskite crystals, the intensity of the Raman lines for the described electron and hole spin flips are proportional to \cite{Rodina2022}:
\begin{equation}\label{Eq1}
    I^{1e(1h)} \propto |{\bm e}^*\times {\bm e}^0|^2 \sin^2 \theta \, .
\end{equation}
Here ${\bm e}^0$ and ${\bm e}$ are the polarization vectors of the incident and scattered light, respectively, and  $\theta$ is the angle between the magnetic field direction and the light propagation direction. For the sake of clarity, we will refer to the SF mechanism described by Eq.~\eqref{Eq1} as the exchange interaction mechanism. 

We remind here that the e-SF and h-SF lines are observed also in the Faraday geometry, see Figs.~\ref{fig:SI5}(a,b). It follows from Eq.~\eqref{Eq1}, that in the exact Faraday geometry ($\mathbf{B}_{\rm F}\parallel \mathbf{k}$ for $\theta=0^\circ$) the spin-flip process is forbidden in cubic-phase perovskites. This situation might be different for perovskite crystals with reduced symmetry, e.g., in our case for CsPbBr$_3$ crystals in the orthorhombic phase. In this case, first of all, the  anisotropy reveals itself in the difference between properties along the main c-axis and in the plane perpendicular to the c-axis, as expected already in the tetragonal phase. Indeed, as demonstrated in the experimental section, we observe a difference between the electron (hole) $g$-factor values for the magnetic field directed along, $g_{e(h)\parallel}$, and perpendicular, $g_{e(h)\perp}$, to the c-axis. An in-plane anisotropy of the $g$-factors, possible in the orthorhombic phase, cannot be distinguished experimentally. Therefore, we analyze below in our theoretical considerations only the symmetry reduction from the cubic to the tetragonal phase. 

The first consequence of this reduced symmetry is the fact that the direction of the effective magnetic field inside the sample, $\tilde {\bm b}_{e(h)} = (g_{e(h)\perp} {\bm b}_\perp + g_{e(h)\parallel}{\bm b}_\parallel)/g_{e(h)}$, declines from the direction of the external magnetic field, ${\bm b}={\bm B}/B$.  One can see, that this declination becomes important when the anisotropic c-axis is tilted with respect to the magnetic field direction ${\bm b}$, however, $\tilde {\bm b}_{e(h)} \parallel {\bm b}$ for $\theta_B = 0^\circ$ and $\theta_B = 90^\circ$. 

The second consequence is the anisotropic splitting of the bright exciton triplet state excited by light. The split states are characterized by the exciton spin projections $F=\pm 1$ and $F=0$ on the c-axis with the anisotropic energy splitting given by $\Delta_{\rm an}$ between them. If $\Delta_{\rm an}$ is larger than the exciton linewidth $\hbar \Gamma$ and the Zeeman splittings $\Delta E_{e(h)}=g_{e(h)}\mu_{\rm B}B$,  the SF intensity provided by the exchange mechanism is given by:
 \begin{equation}\label{Eq1c}
    I^{1e(1h)} \propto   \frac{g_{e(h)\perp}^2}{g_{e(h)}^2} |[{\bm e}^*\times
 {\bm e}^0] \cdot {\bm c}|^2 \sin^2 \theta_B \, .
\end{equation}
Here,  $(g_{e(h)\perp}/g_{e(h)})\sin \theta_B = \sin \theta_{\tilde b}$, where $\theta_{\tilde b}$ is the angle between the magnetic field direction $\bm {\tilde b}$ in the sample and the c-axis. Note that Eq.~\eqref{Eq1c} is similar to that describing the e-SF in quasi-two-dimensional CdSe nanoplatelets \cite{Rodina2020,Kudlacik2019}. Therefore, Eq.~\eqref{Eq1c} is also valid for two-dimensional perovskites or quasi-two-dimensional perovskite nanoplatelets. 

In contrast to the cubic case described by Eq.~\eqref{Eq1}, the intensity of the single SF line described by Eq.~\eqref{Eq1c} depends not only on the angle $\theta$ between the light wave vector $\bm k$ and the magnetic field direction. First, it depends on the angle $\theta_B $ between the anisotropic c-axis and the magnetic field direction. Second, it  depends on the angle $\theta_k$ between the c-axis and the light wave vector $\bm k$. Indeed, if the light excites only the exciton states with projections $\pm 1$ on the c-axis, one obtains $|[{\bm e}^*\times  {\bm e}^0] \cdot {\bm c}|^2 = |{\bm e}^*\times  {\bm e}^0|^2 \cos^2 \theta_k$. In this case, the single SF is forbidden in our experimental geometry with ${\bm k} \perp {\bm c}$ both in the Faraday and Voigt configurations because of $\cos^2 \theta_k = 0$. As the angle between the ${\bm k}$ vector and the c-axis is not exactly $90^\circ$ in the experimental setup (see Methods), the SF becomes allowed both in the Faraday (${\bm B}_{\rm F} \parallel {\bm k}$) and Voigt (${\bm B}_{\rm V} \perp {\bm k}$) geometries. However, the SF described by Eq.~\eqref{Eq1c}  is forbidden in the strict case of ${\bm B} \parallel {\bm c}$ because of $\sin^2 \theta_B = 0$. This contradicts to our experimental observations, see  Figs.~\ref{fig:SI5}(a,b). Note that similar spectra are observed in all experimental geometries shown schematically in Fig.~\ref{fig:SI8}. Therefore, we conclude that the anisotropic exciton splitting $\Delta_{\rm an}$ is negligible in our sample and will continue to analyze the single SF with the help of  Eq.~\eqref{Eq1} assuming that it becomes allowed in our Faraday geometry due to $\theta \approx 10^\circ$ (see Methods).

Let us analyze the polarization selection rules for the single SF predicted by Eq.~\eqref{Eq1}. For linearly polarized light, cross-polarization is expected to be favorable. One can see in Fig.~\ref{fig:S2}(a), that in the Voigt geometry the e-SF and h-SF lines are indeed much stronger for the HV polarization compared to the HH one. 

For circularly polarized light, Eq.~\eqref{Eq1} predicts co-circular polarization selection rules. One can see in Figs.~\ref{fig:SI5}(a,b) that indeed the e-SF and h-SF lines 
are stronger for the $\sigma^+\sigma^+$ than for the $\sigma^+\sigma^-$ polarizations in the anti-Stokes spectral range and for the $\sigma^-\sigma^-$ than for the $\sigma^-\sigma^+$ polarizations in the Stokes range. However, the signals are nearly unpolarized for $\sigma^+$ excitation in the Stokes case and for $\sigma^-$ excitation in the anti-Stokes case. In other words, there is a violation of the circular-polarization selection rules, which depends on the polarization of the excitation light and the initial spin state of the resident carrier. 

Such phenomenology points out to the existence of an additional SF mechanism for a specific spin-alignment of the intermediate states. Such a mechanism could be the Larmor precession of the unpaired spin in the singlet trion state. Indeed, Eq.~\eqref{Eq1} is derived for the case when $\Delta E_{e(h)} = g_{e(h)}\mu_{\rm B}B  \ll \hbar \Gamma$. This condition allows one to neglect the exciton Zeeman splitting, the mixing of exciton states by the magnetic field, as well as any effects related to the Larmor precession of the photoexcited carriers during their coherence time. If the condition $g_{e(h)}\mu_{\rm B}B \ll \hbar \Gamma$  is not fulfilled, the Larmor precession of the unpaired spin 1/2 might compete with the exchange interaction between the resident and the photoexcited carrier~\cite{Rodina2022}. 

The spin level structure of the e-SF intermediate states excited with circularly polarized light are shown in Figs.~\ref{fig:SI5}(c-f).  In the Stokes process, when the initial resident electron is in the spin-down state, the negatively charged singlet T$^-$ trion  can be formed under excitation with $\sigma^+$ polarized light, see Fig.~\ref{fig:SI5}(c). In this singlet trion state, the spins of the two electrons are antiparallel, while the spin of the hole is unpaired. We show the hole spin as a superposition of spin-up and spin-down states with coefficients $\alpha$ and $\beta$. Right after excitation it has spin up with respect to the ${\bm k}$ direction ($\alpha=1, \beta= 0$). The fast Larmor precession of the unpaired hole in the magnetic field, which is not strictly parallel to ${\bm k}$, during the coherence time creates the mixed spin state with $\beta \ne 0$. As a result, the circular polarization of the scattered light might be $\sigma^-$, which means cross-polarized to the excitation. The exchange-mediated flip-stop mechanism with co-polarized $\sigma^+$ scattering also remains possible. As a result, the SF via the singlet trion intermediate state follows no strict polarization rules. 

In contrast, excitation with $\sigma^-$ polarized light generates electron and hole with spin down states and the probability to form a singlet trion in the Stokes process is very low, see  Fig.~\ref{fig:SI5}(d). In this case the only possible mechanism is admixture of the resident electron with opposite spin ($\beta \ne 0$) due to the exchange interaction mechanism described above. Hence, the polarization selection rules remain more strict in the Stokes case. 

In turn, the inital states of the resident electron and the hole with spin up orientation can be involved in the anti-Stokes process, see Figs.~\ref{fig:SI5}(e,f). Here, the singlet trion formation and violation of the selection rules is expected for $\sigma^-$ excitation.

We discuss now briefly the double electron SF line with the Raman shift described by twice the electron $g$ factor $2g_e$. Such a line was originally found in moderately doped n-CdS~\cite{Scott1972,Economou1972}. It was assigned to a spin-flip of a pair of electrons bound at two neighboring donors, both interacting with a photogenerated localized exciton. Recently, we found such a process experimentally and considered it theoretically for CdSe colloidal nanoplatelets~\cite{Kudlacik2019,Rodina2020}, where two localized electrons were responsible for the double SF. We suggest that in case of CsPbBr$_3$, the process also involves two localized electrons, which wave functions overlap with the same exciton. A theoretical model of the double electon and double hole, 2e-SF and 2h-SF, processes as well as the combined ${\rm(e\pm h)}$-SF in cubic phase perovskites was developed in Ref.~\onlinecite{Rodina2022}. The processes responsible for the 2e-SF signals are shown schematically in Fig.~\ref{fig:S2}(d) for the Stokes components in the Voigt geometry. Note that the double flip-stop exchange mediated SF is co-polarized with linearly polarized light. The absence of an offset in the extrapolation of the 2e line shift to zero magnetic field (see Fig.~\ref{fig:1}(c)) evidences that the involved resident electrons have a small exchange interaction (not exceeding 5~$\mu$eV) with each other, i.e., their wave functions overlap very weakly~\cite{Kudlacik2019,Rodina2020}.

\subsection{Exciton spin-flip Raman scattering} 
\label{exc}

\begin{figure}
\centering
\includegraphics[width=0.5\textwidth]{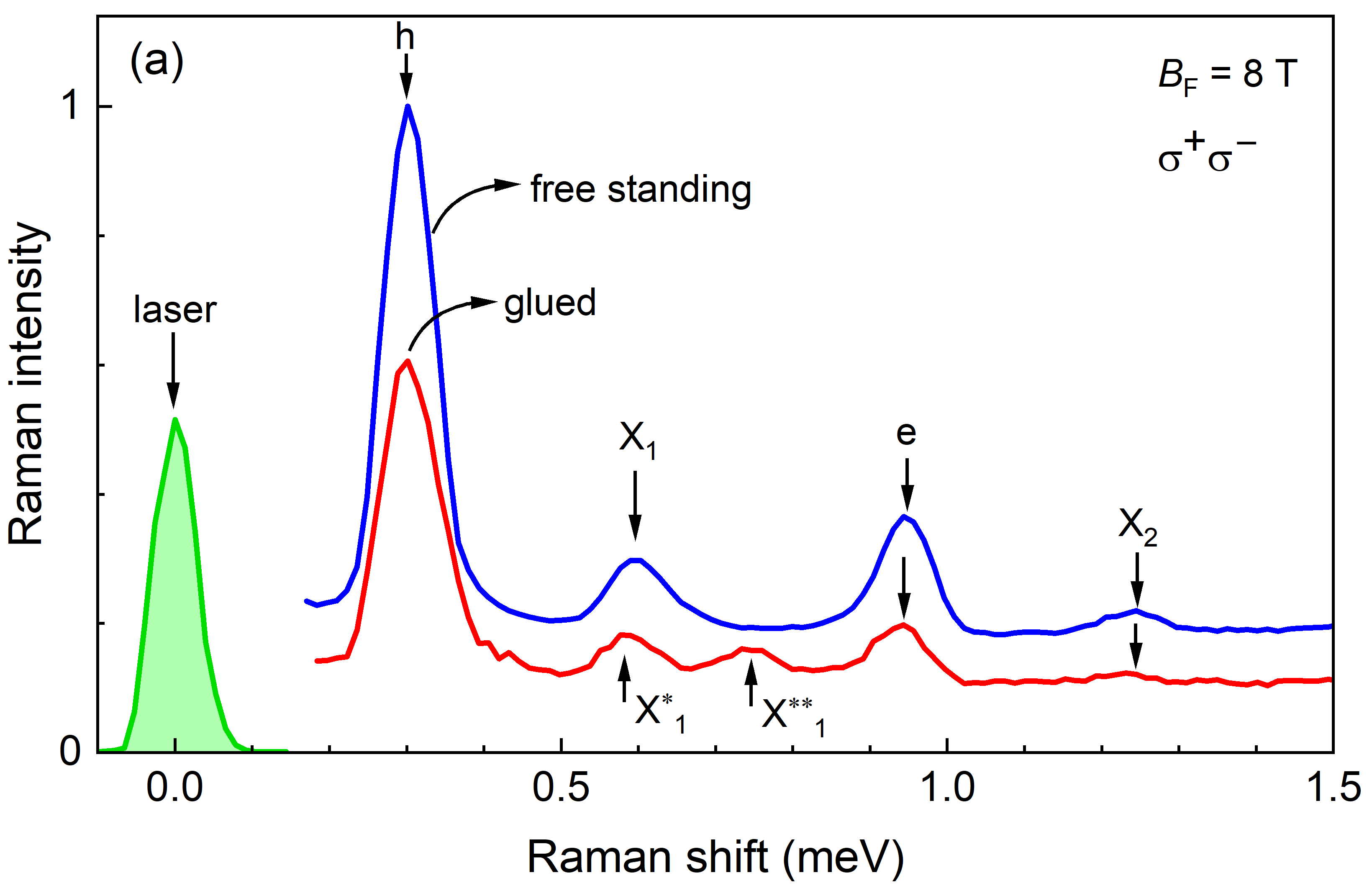}
\includegraphics[width=0.5\textwidth]{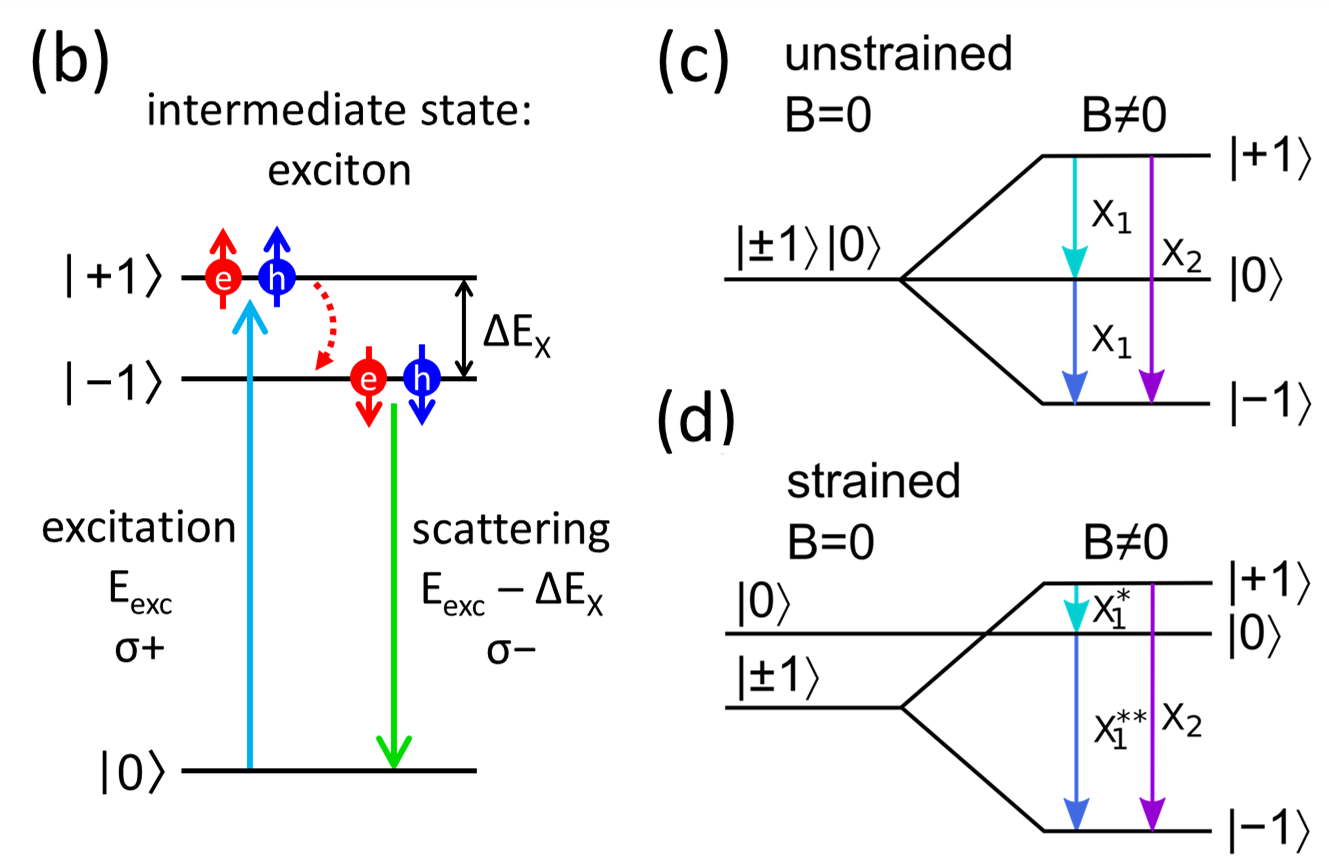}
\caption{{Strain effect on exciton spin-flip. }  
({a}) SFRS spectra of a CsPbBr$_3$ crystal measured in cross-circular polarizations at $B_{\rm F}=8$~T applied in the Faraday geometry ($\mathbf{B}_{\rm F} \parallel \mathbf{k}$,  $\mathbf{B}_{\rm F} \perp \mathbf{c}$). The blue spectrum is recorded for strain-free sample mounting, and the red one is for the sample glued on a copper sample holder. $T=1.6$~K.  ({b}) Scheme of the exciton spin-flip process with changing the spin by $\pm 1$ via the emission of an acoustic phonon. The blue arrow indicates $\sigma^{+}$ excitation and the green arrow $\sigma^{-}$ scattering.   ({c, d}) Schematic illustration of the spin-flip transitions on the exciton without and with removing the degeneracy of the $\pm 1$ and $0$ states by strain. 
}
\label{fig:2}
\end{figure}

Let us turn now to the X$_1$ and X$_2$ lines in the Raman spectra, which we associate with spin-flip of the exciton itself. These lines are seen both in the Faraday geometry (Figs.~\ref{fig:SI5}(a), \ref{fig:SI5}(b)  and \ref{fig:2}(a)), where they are stronger in cross-circular polarization, and in the Voigt geometry, where they are stronger in the co-linear polarization (Fig.~\ref{fig:S2}(a)). The Zeeman splitting of the X$_2$ line in the Faraday geometry corresponds to $g_{X2\perp}=2.72$ (Fig.~\ref{fig:1}(c)), which exactly matches the relation $g_X=g_e+g_h$ for the bright exciton $g$-factor in lead halide perovskites~\cite{kopteva2023_gX}. This allows us to assign the X$_2$ line to the spin-flip of the bright exciton between the $|+1\rangle$ and $|-1\rangle$ states with the projections $F=+1$ and $F=-1$ on the magnetic field direction, respectively. The scheme of this process is shown for the Stokes range in Fig.~\ref{fig:2}(b). The exciton in the $|+1\rangle$ state is created by $\sigma^+$ excitation. The spin-flip to the $|-1\rangle$ state is mediated by the electron-hole exchange interaction of the exciton, provided by the electron spin-flip, as acoustic phonons couple electrons from the $\Gamma_7$ and $\Gamma_8$ bands~\cite{Rodina2024}. It does not involve any resident carriers, but instead is  accompanied by the emission (in Stokes) or absorption (in anti-Stokes) of an acoustic phonon with resonant energy $E_{\rm ph}=\Delta E_X = g_X \mu_{\rm B} B$. The  scattered photon has circular polarization opposite to the incident one, i.e., $\sigma^-$ polarization. Therefore, the exciton spin-flip is observable in crossed polarization ($\sigma^+\sigma^-$ in Stokes or $\sigma^-\sigma^+$ in anti-Stokes). In this process, the photoexcited exciton is assumed to be localized and not form an exciton-polariton, which we will consider in Section~\ref{brill}.

The X$_1$ line has about a twice smaller $g$-factor ($g_{X1\perp}=1.30$) compared to $g_{X2\perp}=2.72$, see Fig.~\ref{fig:1}(c).  This allows us to assign this line to the exciton spin-flip transitions $|+1\rangle \to |0\rangle$ and $|0\rangle \to |-1\rangle$, see Fig.~\ref{fig:2}(c). We remind here that up to now we did not observe any evidence for the exciton anisotropic Zeeman splitting caused by the presence of the c-axis and, therefore, consider a three-fold-degenerated bright exciton state. Such transitions mediated by the electron-hole exchange and electron-phonon interaction are considered theoretically in Ref.~\onlinecite{Rodina2024} for the cubic phase perovskites. For further experimental confirmation of this interpretation we use the fact that the spin-flip Raman process involving the exciton spin structure is strongly sensitive to sample strain and deformation~\cite{BirPikus}. 

Figure~\ref{fig:2}(a) shows SFRS spectra of the CsPbBr$_3$ crystal measured in cross-circular polarizations at $B_{\rm F}=8$~T applied in the Faraday geometry. The blue spectrum is for strain-free sample mounting, and the red one is for the sample glued on a copper sample holder. The glued sample is slightly stressed due to different coefficients of the thermal expansion of the copper and the sample. This local uncontrolled strain partially removes the three-fold degeneracy of the exciton ground state, leading to its anisotropic splitting into the $F=0$ and $\pm1$ sublevels, see Fig.~\ref{fig:2}(d). Here the projections are done relative to the strain axis.  When the direction of the magnetic field is not coinciding with a strain axis, it mixes the states allowing excitation and recombination of all three exciton states. Additionally, it activates the exciton spin-flip processes $|+1\rangle \to |0\rangle$ and $|0\rangle \to |-1\rangle$, labeled as X$^*_1$ and X$_1^{**}$, respectively. For the strain-free mounted sample the Raman shifts of the X$_1^*$ and X$_1^{**}$ lines are equal to each other and carry the X$_1$ label (see the blue spectrum in Fig.~\ref{fig:2}(a) and the scheme in Fig.~\ref{fig:2}(c)) and amount to half of that of the X$_2$ line. The energy shift between the X$_1^*$ and X$_1^{**}$ lines depends on the applied strain and amounts to 0.17~meV in the studied sample, see the red spectrum in Fig.~\ref{fig:2}(a). 

\subsection{Comparison of the exciton and combined e-h spin flip mechanisms.}
\label{mecanism_X}

In Section~\ref{exc} we described the observed X$_2$ and X$_1$ lines as processes with change of the photoexcited exciton spin by $\pm 2$ (with the Raman shift $\pm \Delta E_X =\pm (g_e+g_h)\mu_{\rm B}B$) or by $\pm 1$ (with the Raman shift $\pm \Delta E_X/2 =\pm (g_e+g_h)\mu_{\rm B}B/2$), respectively. Figure \ref{fig:2} illustrates the Stokes parts of these processes.  As shown in Ref.~\onlinecite{Rodina2024}, in this case both exciton spin flips are mediated by two perturbations: (1) the electron-hole exchange interaction,  which couples the ground exciton states to the excited states of the electron-hole pairs with the same total spin projection, and (2) the electron-acoustic phonon interaction, which changes the electron spin. Thus, the initial and final states for these exciton spin-flips do not assume the presence of resident carriers or any preceding excitation of the sample. However, they involve the emission or absorption of an acoustic phonon. The phonon energy should be equal to the respective Raman shift. 

\begin{figure}
\centering
\includegraphics[width=0.5\textwidth]{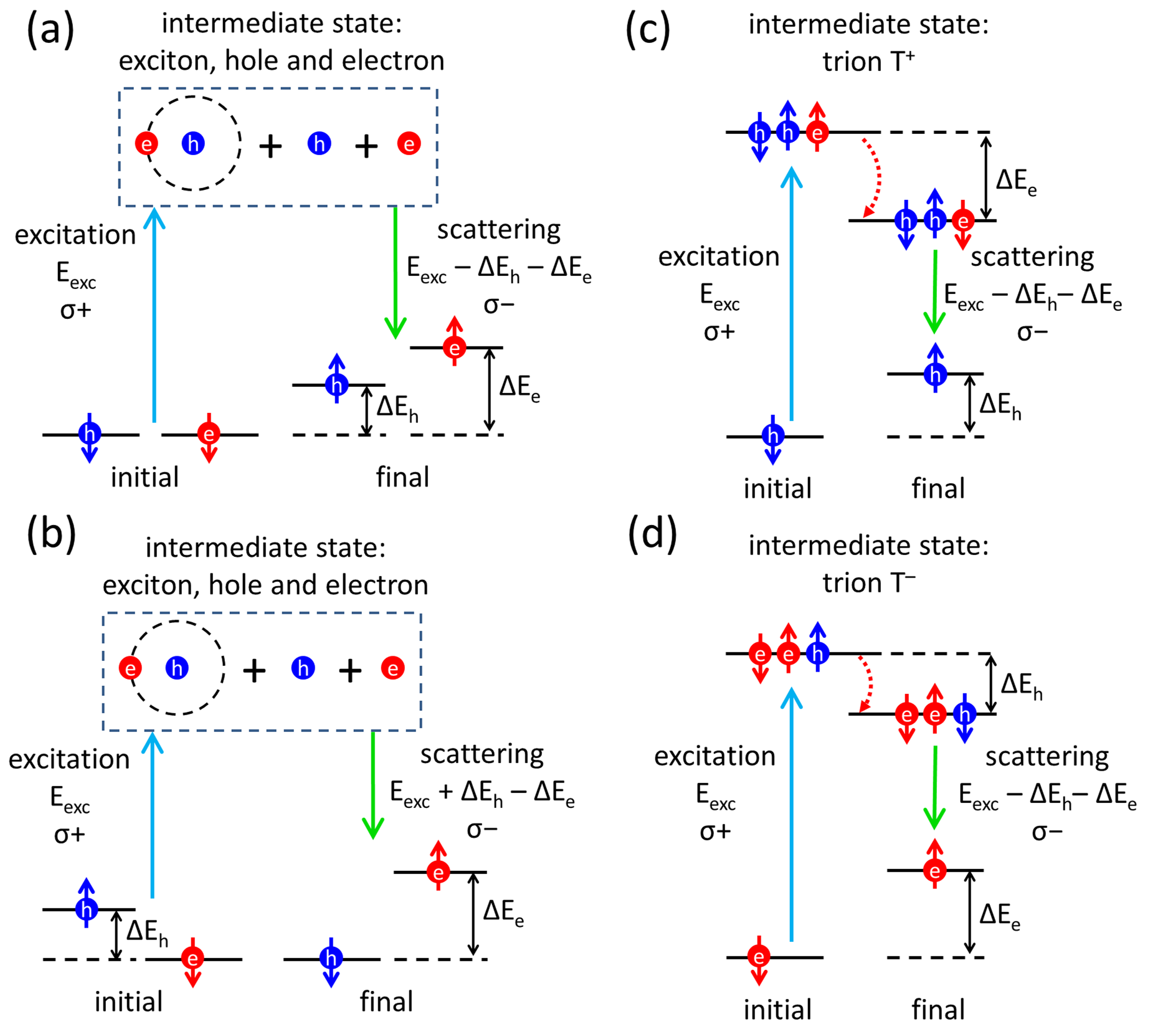}
\caption{{Alternative mechanisms of SFRS involving an exciton Raman shift (Stokes process).}  
({a, b}) Scheme of the combined electron and hole spin flips: e+h SF (a) and e-h SF (b). 
({c, d}) Alternative schemes for the $\Delta E_e+\Delta E_h$ Raman shift via a positive (c) and anegative (d) trion intermediate state. The spin-flip of the electron (c) or hole (d) in the intermediate state is assisted by the emission of a resonant acoustic phonon. In the schemes, the blue arrows stand for $\sigma^{+}$ excitation, while the green arrows indicate $\sigma^{-}$ scattering. Note that for the e-h SF (b), both $\sigma^{+}\sigma^{-}$ and $\sigma^{-}\sigma^{+}$ scattering are possible in the Stokes range. 
}
\label{fig:2b}
\end{figure}

Importantly, there is another possibility to describe both the X$_2$ and X$_1$ lines, also considering the exciton spin flip by $\pm 2$ and $\pm 1$, but without participation of any phonon. In this case, the initial and final states assume the presence of the already excited bright exciton and the intermediate state -- resulting in the formation of a singlet biexciton. Such process suggested in Ref.~\onlinecite{Rodina2022} is characterized by the same Raman shifts and the same polarization rules for the X$_2$ (cross circular polarization) and X$_1$ lines as the phonon-assisted spin-flip. Thus, the phonon-assisted and biexciton-mediated exciton spin-flips are nearly indistinguishable. However, the biexciton formation is much less probable using not too high excitation powers. 

Next, we discuss another possible spin-flip process that results in the same Raman shift $\Delta E_X =(g_e+g_h)\mu_{\rm B}B =\Delta E_e + \Delta E_h$ and has the same polarization selection rules (e.g., $\sigma^{+} \sigma^{-}$ in Stokes) as the X$_2$ spin-flip process in  Fig.~\ref{fig:2}(b). We call this process a combined e+h SF, as it assumes the participation of one resident electron and one resident hole in the initial and final states. In contrast to the biexciton-mediated SF~\cite{Rodina2022} described above, the resident electron and the resident hole are localized separately from each other. The scheme in Fig.~\ref{fig:2b}(a) shows an example of such a Stokes process with electron and hole in spin down initial states. The exciton is excited by $\sigma^+$ polarized light and interacts with the electron and the hole in the intermediate state. The scattered light has $\sigma^-$ polarization with the photon energy given by $E_{\rm exc}-\Delta E_h - \Delta E_e$. In the final state, both the resident electron and the resident hole have spin up. The polarization selection rules (cross circular polarization) and the Stokes shift $\Delta E_e + \Delta E_h = \Delta E_X$ are the same as expected for the phonon-assisted exciton spin-flip shown in Fig.~\ref{fig:2}(b) or the biexciton-mediated exciton SF.  However, when the electron and hole are not bound to an exciton in the initial and final states, an additional SF with energy shift $\Delta E_X/2$ is not predicted. Instead, along with the e+h SF, a new combined e-h SF process is predicted. It should be also observed with the Raman shift given by $(g_e-g_h)\mu_{\rm B}B$, see Fig.~\ref{fig:2b}(b). In our case, just by incidence, $(g_e-g_h) \approx (g_e+g_h)/2 = \Delta E_X/2$ corresponding to the X$_1$  line. But the shift of X$_1$ with strain proves that the process corresponds to the exciton spin-flip and not the one involving the combined electron and hole.

In addition, one can suggest two more combined SF processes, which can describe the X$_2$ line, but fail to explain the X$_1$ line. These processes assume the presence of one resident carrier, either an electron or a hole, in the initial and final states. The respective Stokes SF schemes are shown in Figs.~\ref{fig:2b}(c) and \ref{fig:2b}(d). Excitation with $\sigma^{+}$ polarized light creates a singlet positive, T$^+$, or negative T$^-$, trion in the intermediate state. Then the spin of the unpaired carrier is flipped with assistance of resonant acoustic phonon emission and the subsequent recombination of the trion leaves behind the resident carrier with opposite spin. Note that these processes have some similarity to the single SF mediated by a singlet trion state, see Figs.~\ref{fig:SI5} (c,f). However, they involve the participation of a resonant phonon and, therefore, have a Raman shift equal to $\Delta E_X$. In contrast to the single SF, these processes are allowed in ideal Faraday geometry. 


\section{Brillouin light scattering on exciton-polaritons}
\label{brill}

\begin{figure*}
\centering
\includegraphics[width=1\textwidth]{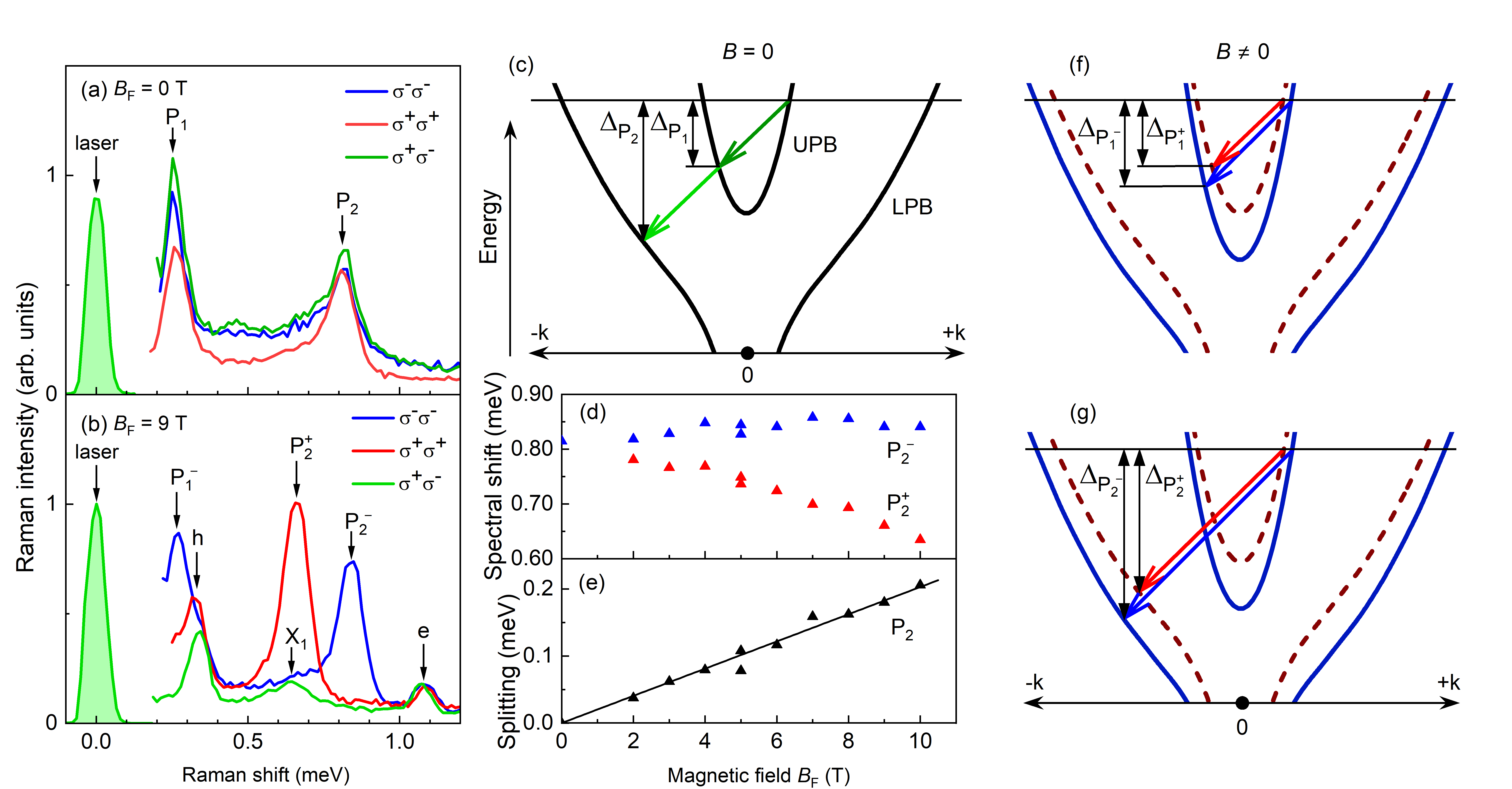}
\caption{{Brillouin light scattering on exciton polaritons.} 
({a}) Light scattering spectra at $B_{\rm F}=0$~T in co-polarization $\sigma^-\sigma^-$ (blue line) and $\sigma^+\sigma^+$ (red line), and in cross-polarization  $\sigma^+\sigma^-$ (green line), $\mathbf{k} \perp \mathbf{c}$. ({b}) Light scattering spectra at $B_{\rm F}=9$~T in co-polarization measured in the Faraday geometry ($\mathbf{B}_{\rm F} \parallel \mathbf{k}$ and $\mathbf{B}_{\rm F} \perp \mathbf{c}$). The green line shows the spectrum in cross-polarization $\sigma^+\sigma^-$, where the P$_1$ and P$_2$ lines are not visible.  ({c}) Scheme of Brillouin light scattering on exciton polaritons at zero magnetic field. ({d}) Magnetic field dependence of the spectral shifts of the P$_2$ line in different co-polarizations. ({e}) Magnetic field dependence of the P$_2$ line splitting. The line is a linear interpolation with the slope of 0.02 meV/T. ({f, g}) Scheme of the spin-dependent Brillouin light scattering on exciton-polaritons in finite magnetic field.   }
\label{fig:3}
\end{figure*}

Figure~\ref{fig:3}(a) shows light scattering spectra in co- and cross-circular polarizations. At $B=0$ there are the two lines P$_1$ and P$_2$ shifted from the laser energy by 0.26 and 0.81~meV, respectively. We assign these lines to resonant Brillouin scattering (RBS) on exciton polaritons, a phenomenon that has been reported for various III-V and II-VI semiconductors~\cite{Ulbrich1977,Koteles1979,Weisbuch1982,Honerlage1985,Sugisaki2001,Sandfort2009}. The origin of the RBS lines is shown schematically in Fig.~\ref{fig:3}(c) for the simplest case of two polariton branches. Here, the upper (UPB) and lower (LPB) exciton-polariton branches are shown. The exciton is photogenerated on the UPB and scatters at a longitudinal or transversal acoustic phonon either within the UPB (P$_1$ line) or onto the LPB (P$_2$ line). The spectral shift of the lines ($\Delta_{\rm P_1}$ or $\Delta_{\rm P_2}$) is determined by the dispersions of the exciton-polaritons and of the acoustic phonons, so that the momentum conservation is fulfilled. Note that at $B=0$ the P$_1$ and P$_2$ lines are well seen in both co- and cross-circular polarizations, see Figs.~\ref{fig:3}(a) and \ref{fig:1}(b). 

In a magnetic field applied in Faraday geometry, even as strong as 9~T, the P$_1$ and P$_2$ lines show a small energy shift (Figs.~\ref{fig:3}(b,d)). Each line splits into two components denoted as P$_1^{+}$, P$_1^{-}$ and P$_2^{+}$, P$_2^{-}$ active either in $(\sigma^+,\sigma^+)$ or $(\sigma^-,\sigma^-)$ polarization.  Note that the P$_1^{+}$ line shifts too close to the laser line and we can not resolve it, therefore, only the energy splitting of the P$_2$ line is plotted in Fig.~\ref{fig:3}(e). The splitting is linear with magnetic field, having a slope of 0.02~meV/T. Note that it is considerably smaller than the Zeeman splitting of the carriers and excitons shown in Fig.~\ref{fig:1}(c).  

In Figures~\ref{fig:3}(f,g) we show schematically the modification of the exciton-polariton dispersion in a magnetic field, accounting for the exciton Zeeman splitting and its coupling to the photons with projections $\pm 1$ on the magnetic field. Each exciton-polariton splits into two branches, which can absorb or emit circularly polarized light with opposite signs. Note that the splitting depends on the value of $|k|$ for each brunch differently: it is larger for the exciton-like states and vanishes for the photon-like states. The corresponding, polarization-dependent Brillouin scattering processes are shown by the arrows. This explains the magnetic field splitting of the P$_2$ line in opposite co-polarizations, as seen in Figs.~\ref{fig:3}(d,e).

It is interesting that the P$_1$ and P$_2$ lines loose their intensity for cross-circular polarization and are not visible at $B_{\rm F}=9$~T, see the green spectrum in Fig.~\ref{fig:3}(b). The absence of the strict polarization selection rules in zero magnetic field and their recovery at $B_{\rm F}=9$~T can be explained as follows. In zero magnetic field, the degeneracy of the exciton-polariton branches can be lifted by the birefringence effect in the considered geometry with ${\bm k} \perp {\bm c}$ because of the refractive index anisotropy. As a result, they interact with circularly polarized light of both signs and the scattering is unpolarized. When the magnetic-field-induced splitting overcomes this anisotropy splitting, the strict selection rules for the exciton-polariton branches are reestablished, so that the scattering with longitudinal acoustic phonons does not change the exciton spin projection.

\section{Conclusions}

In this study we have demonstrated that the Raman and Brillouin light scattering techniques are suitable and informative tools for investigating the spin-dependent properties of excitons and charge carriers in lead halide perovskite semiconductors. This has been exemplified for CsPbBr$_3$ crystals for which in strong magnetic fields up to 10~T a broad variety of spin-flip lines has been observed in the vicinity of the laser line, which is resonant with the exciton-polariton energy. From the Raman shifts of the electron, hole, and exciton SF lines, their $g$-factor values and their anisotropies have been evaluated. The selection rules for the spin-flip Raman scattering processes for excitons and for resident carriers interacting with the excitons have been analyzed theoretically for perovskite crystals with an anisotropic c-axis, which is inherent to the orthorhombic crystal symmetry. We have considered several mechanisms for the combined spin-flip Raman scattering processes involving resident carriers and photoexcited excitons, suggested new mechanisms involving trions in the intermediate scattering state, and discussed their applicability to the gained experimental data. A double spin-flip process of the electrons has been found and explained by the interaction of the photogenerated exciton with two resident electrons. We have shown that the spin-dependent Raman light scattering is a sensitive optical tool for local strain.
The observation of resonant Brillouin scattering lines on the exciton-polariton branches and their splitting in a magnetic field opens the possibility for the studying the exciton-polariton dispersions and their modification by an external magnetic field. These experimental results and the evaluated material parameters, extended to several lead halide perovskite materials,  will provide an in-depth understanding of their spin-dependent optical properties. They can be also used for testing and refining theoretical approaches for calculating the band structure parameters.

\subsection{Acknowledgements}
The authors are thankful to E. L. Ivchenko, M. A. Semina, and M. M. Glazov for fruitful discussions. We thank G. S. Dimitriev for technical support of the measurements. The authors acknowledge financial support by the Deutsche Forschungsgemeinschaft via the SPP2196 Priority Program (Project YA 65/28-1, No. 527080192). The work of A.V.R. on the theoretical analysis was supported by the Russian Science Foundation (Grant No. 23-12-00300).



\subsection{ORCID}

Ina V. Kalitukha:  0000-0003-2153-6667  \\  
Victor F. Sapega:  0000-0003-3944-7443  \\  
Dmitri R. Yakovlev: 0000-0001-7349-2745  \\  
Dennis Kudlacik:  0000-0001-5473-8383  \\  
Damien Canneson:  0009-0001-1144-6900  \\ 
Yury G. Kusrayev:   0000-0002-3988-6406  \\  
Anna V. Rodina:   0000-0002-8581-8340 \\  
Manfred Bayer:  0000-0002-0893-5949  \\   


\clearpage

\setcounter{equation}{0}
\setcounter{figure}{0}
\setcounter{table}{0}
\setcounter{page}{1}
\renewcommand{\theequation}{S\arabic{equation}}
\renewcommand{\thefigure}{S\arabic{figure}}
\renewcommand{\thepage}{S\arabic{page}}
\renewcommand{\thetable}{S\arabic{table}}

\begin{center}

\section*{Supporting Information}

\textbf{\large Spin-dependent Raman and Brillouin light scattering on excitons in CsPbBr$_3$ perovskite crystals}\\
{I. V. Kalitukha, V. F. Sapega, D. R. Yakovlev, D. Kudlacik, D. Canneson, Yu. G. Kusrayev, A. V. Rodina,  and M. Bayer}
\end{center}

\subsection*{S1. Sample}
\label{sec:S1}

The studied rectangular CsPbBr$_3$ sample is crystallized in the orthorhombic modification. The crystal has one selected (long) direction along the c-axis [001] and two nearly identical directions along the [\=110] and [110] axes.
The crystal has an elongated shape with sizes of $5\times 2 \times 2$~mm$^3$. 

\begin{figure}[h]
\centering
\includegraphics[width=0.3\textwidth]{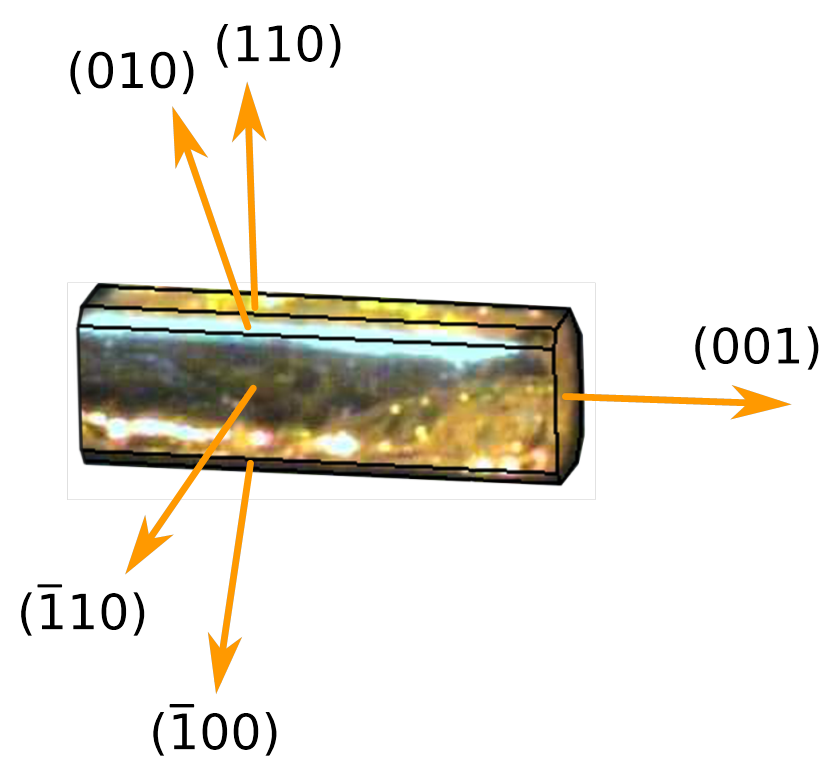}
\caption{Image of the CsPbBr$_3$ sample with the crystals axes indicated. (001) corresponds to the c-axis.}
\label{fig:SI1}
\end{figure}

\subsection*{S2. Experimental geometries}
\label{sec:S2}

\begin{figure}[h]
\centering
\includegraphics[width=0.5\textwidth]{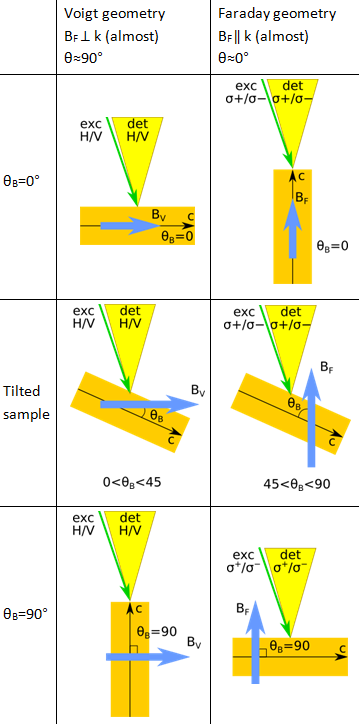}
\caption{Experimental geometries.}
\label{fig:SI8}
\end{figure}

Figure~\ref{fig:SI8} shows schemes of the experimental geometries illustrating the relationship between the wave vector $\mathbf{k}$, the magnetic field $\mathbf{B}$, and the crystal c-axis using the angles $\theta$ (angle between $\mathbf{B}$ and $\mathbf{k}$) and $\theta_B$ (angle between $\mathbf{B}$ and $\mathbf{c}$).

Raman spectra were measured in co- and cross-linear polarizations in the Voigt geometry ($\mathbf{B}_{\rm V} \perp \mathbf{k}$) and in co- and cross-circular polarizations in the Faraday geometry ($\mathbf{B}_{\rm F} \parallel \mathbf{k}$), corresponding to the right and left columns in Fig.~\ref{fig:SI8}).

In order to measure the different components of the $g$-factor tensor, the angle $\theta_B$ between the magnetic field $\textbf{B}$ and the c-axis was varied in the range $0^{\circ} \le \theta_B \le 90^{\circ}$ by rotating the sample. During this process, the angle between the magnetic field and the incident laser remained the same for both geometries, namely $\mathbf{B}_{\rm V} \perp \mathbf{k}$ for the Voigt geometry while tilting the sample ($0^{\circ} \le \theta_B \le 45^{\circ}$) and $\mathbf{B}_{\rm F} \parallel \mathbf{k}$for the Faraday geometry while tilting the sample ($45^{\circ} \le \theta_B \le 90^{\circ}$).

\subsection*{S3. SFRS in circular polarization}
\label{sec:S3}

\begin{figure*}[hbt]
\centering
\includegraphics[width=0.9\textwidth]{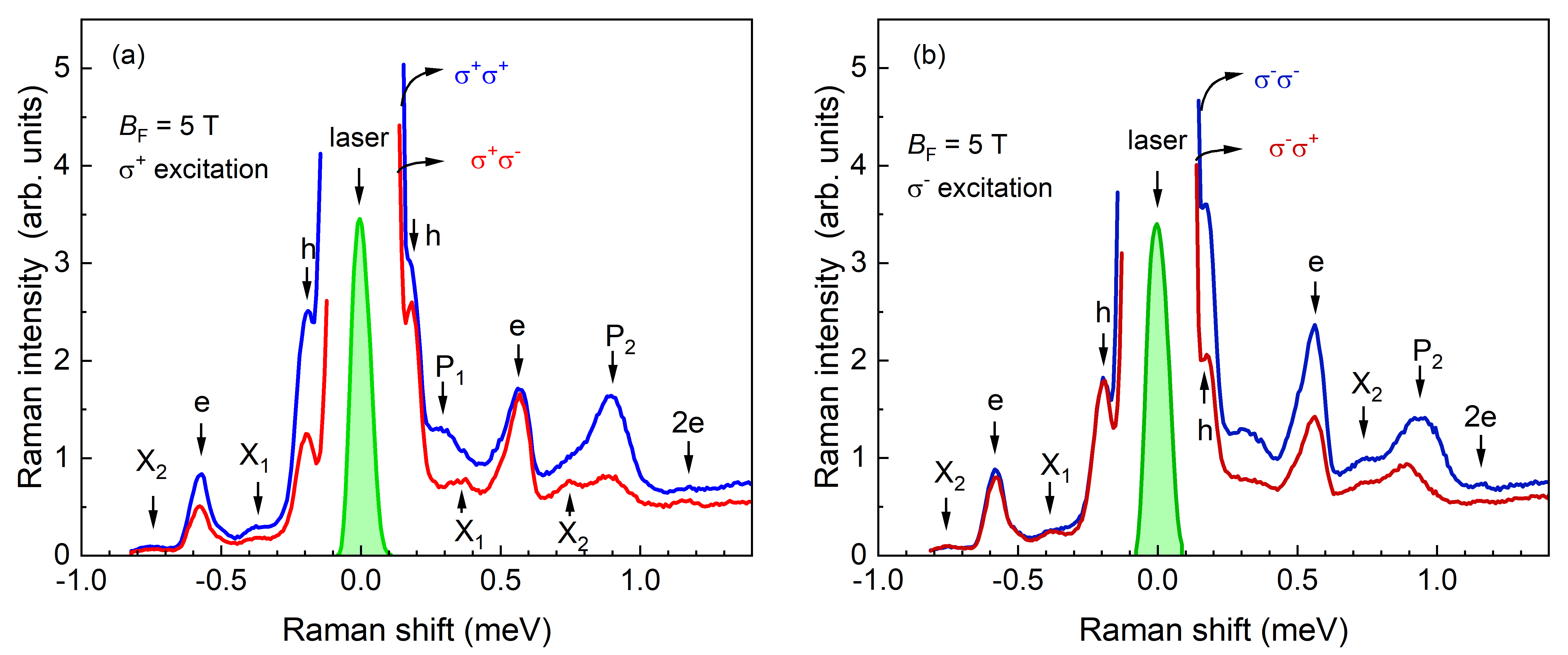}
\caption{Raman spectra in the Faraday geometry for all four configurations of excitation-detection circular polarization. The magnetic field is oriented perpendicular to the sample $c$-axis, $\theta_B = 90^\circ$. The polarization properties almost coincide with the case of a Faraday magnetic field parallel to the crystal c-axis (see Fig.~\ref{fig:SI5}), but the Raman shifts of the electron and hole spin-flip lines are slightly different due to the $g$-factor anisotropy and the Brillouin scattering is suppressed.}
\label{fig:SI4}
\end{figure*}

Figure~\ref{fig:SI4} shows Raman spectra for all combinations of the excitation and detection circular polarization in the Faraday geometry for $\mathbf{B}_{\rm F} \perp \mathbf{c}$, $\theta_B = 90^\circ$.

\end{document}